\documentclass[twoside]{article}
\usepackage[latin1]{inputenc}
\usepackage{t1enc}
\usepackage{a4}
\usepackage{tabularx}
\usepackage{epsf}
\usepackage{graphicx}

\def\bref{\vspace{4pt}\noindent\hangindent=10mm}

\begin{document}

\setcounter{figure}{0}
\setcounter{section}{0}
\setcounter{equation}{0}

\begin{center}
{\Large\bf
Modeling the High-z Universe:\\[0.2cm] 
Probing Galaxy Formation}\\[0.7cm]

Sadegh Khochfar \\[0.17cm]
Max-Planck-Institute for Extraterrestrial Physics \\
Giessenbachstrasse, D-85748 Garching, Germany \\
sadeghk@mpe.mpg.de, http://www.mpe.mpg.de/~sadeghk/
\end{center}

\vspace{0.5cm}

\begin{abstract}
\noindent{\it
We discuss how the conditions at high redshift differ from those at low redshift, and what the impact is on the galaxy population. We focus in particular on the role of gaseous dissipation and its impact on sustaining high star formation rates as well as on driving star-bursts in mergers. Gas accretion onto galaxies at high redshifts occurs on a halo dynamical time allowing for very efficiently sustained star formation. In addition cold accretion flows are able to drive turbulence in high redshift disks at the level observed if at least $20\%$ of the accretion energy is converted into random motion in the gaseous disk. In general we find that the fraction of gas involved in galaxy mergers is a strong function of time and increases with redshift. A model combining the role of dissipation during mergers and continued infall of satellite galaxies allows to reproduce the observed size-evolution of early-type galaxies with redshift. Furthermore we investigate how the evolution of the faint-end of the luminosity function can be explained in terms of the evolution of the underlying dark matter evolution.  }
\end{abstract}

\section{Introduction}
There is ample evidence that the conditions under which galaxies formed at high redshift where quite different to those at later times during the evolution of the universe. Impressive evidence supporting such a view has been lately collected  from detailed integral field spectroscopy of $z \sim 2$ galaxies (F\"orster Schreiber et al. 2006, Genzel et al. 2006). These observations reveal massive, high star forming galaxies that show rotational structure that resemble those of disk galaxies. In contrast to low redshift disk galaxies however, the ratio of gas rotational velocity to velocity dispersion is only of the order few compared to $V / \sigma \sim 10$ at low z. Furthermore it has been well established by now, that the star formation rate in the universe is a declining function of time (Hopkins 2004) and that this is driven by a {\it down-sizing} in the average star formation rate of galaxies of a given mass as a function of time (Juneau et al 2005, Noeske et al. 2007). Besides star formation, merger rates (e.g. Le F\`evre et al 2000) and AGN activity (Hasinger et al. 2007) show an upward trend with redshift, peaking around $z \sim 3-4$ indicating the importance of the high-z universe for the assembly and formation of galaxies. That indeed a large fraction of the stellar mass of massive galaxies is already in place at $z \ge 2$ finds it support from studies of their stellar population (Thomas et al. 2005) and the evolution of the mass density in units of the $z=0$ mass density (P\'erez-Gonz\'alez et al. 2008).

The hierarchical $\Lambda$CDM paradigm has been successful in predicting and reproducing several of the above mentioned trends. In particular the increasing merger fraction as a function of redshift is a natural outcome (kb2001, ks2008, farouhki ma) of this model and driven by the merging history of the underlying dark matter halos. With respect to the evolution of baryonic physics within such evolving haloes early work by (Binney 1977, Silk 1977, Rees \& Ostriker 1977) laid out the ground work. Radiative cooling and the collapse into a rotationally supported disk of initial hot gas at the halos virial temperature, followed  by star formation in this disk where the main mechanism to build up galaxies. Besides its simplifying assumption, e.g. about spherical symmetry of the halo, its profile and other properties of the dark matter and associated gas infall, these models proved to be very successful in predicting galaxy properties and in particular a transition  mass scale at which galaxy formation becomes very inefficient due to long radiative cooling times of the gas (Dekel \& Birnboim 2006), thus causing star formation to slow down or even halt. In comparison to detailed high resolution numerical simulations these models have shown to predict cooling rates too low by a factor of a few, especially in low mass halos and at early times (Keres et al 2005). This discrepancy can be attributed partly to the geometry of the accretion process in cosmological simulations which is not necessarily spherical symmetric as assumed in simplified models, but follows the cosmic filaments. 
However, the essence of the accretion process of cold gas at high redshift can be summarized by fast and efficient, on the host halos dynamical time. In effect cold gas is accreted with a rate comparable to the dark matter accretion rate times the cosmological baryon fraction $f_b \dot{M}_{DM}$. This situation will not hold in massive halos however, where stable shocks heat the infalling gas to the halos virial temperature and cooling times are longer than the dynamical time, and neither at late times when already a substantial fraction of baryons have made it into stars. In the following section we will investigate the impact this efficient provision of cold gas has on the evolution and formation of galaxies.

\section{The Model}
We use the semi-analytic model {\it Simurgh} to model the formation and evolution of galaxies. The dark matter history is calculated using the merger tree proposed by {Somerville et al. 1999} with a mass resolution of $2 \times 10^9 M_{\odot}$. The baryonic 
physics within these dark matter haloes is calculated following recipes 
presented in Springel et al. (2001) and references therein, including a model for the reionizing background 
by {Somerville 2002}. In our simulation, we assume that elliptical galaxies 
form whenever a major merger ($M_1 /M_2 \leq 3.5$ with $M_1 \geq M_2$) takes 
place. We assume that during this process, all the cold gas in the
 progenitor disks will be consumed in  a central starburst, adding to the 
spheroid mass, and that all stars in the progenitor disks will  
contribute to the spheroid as well. Furthermore, we also add the stars of satellite galaxies involved in  minor mergers to the spheroid. The merger time scale for galaxies is calculated using the dynamical friction prescription in Springel et al. (2001) and we find that the predicted merger rate is in good agreement with observations {Khochfar \& Burkert 2001}.
For more modeling details, we refer the reader to Khochfar \& Burkert 2005 and Khochfar \& Silk (2006a). Please note 
that our simulation does not include  
AGN-feedback (Schawinski et al. 2006) 
 or environmental effects (Khochfar \& Ostriker 2007)  that have 
influence on the most massive galaxies. Throughout this paper, we use the following set of cosmological parameters derived from a combination of the 5-year WMAP data with Type Ia supernovae and measurements of baryon acoustic oscillations (Komatsu et al. 2008):
$\Omega_0=0.28$, $\Omega_{\Lambda}=0.72$, $\Omega_b/\Omega_0=0.16$, 
$\sigma_8=0.8$ and $h=0.7$.

\subsection{Evolution of the faint-end luminosity function}
The faint-end of the galaxy luminosity function offers a strong constraint on the efficiency of galaxy formation within small mass dark matter halos. Cosmological N-body simulations predict a power-law slope of $\alpha \sim -2$ at all redshift. Observations of the faint-end luminosity function however, show a slope that is less steep and varies depending on the observed band and environment between $\alpha \sim -0.9$ to $\alpha \sim  - 1.5$ (Ryan et al. 2007). The widely accepted explanation is that feedback from supernovae is hindering star formation in low mass haloes. The rational behind this argument is that the specific energy of hot gas is lower in low mass halos and therefor the amount of cold gas reheated per supernovae is larger than in massive halos (Dekel \& Silk 1986). 
The way that supernovae operate per solar mass stars formed is independent of redshift considering the same stellar initial mass function, thus it comes as a surprise that observations with the Hubble Space Telescope a steepening of the slope with redshift (Ryan et al. 2007). 

In Fig. 1, we show the predicted evolution of $\alpha (z) $ for our 
best-fit local model, i.e. a model that is chosen to best fit the local luminosity function (Khochfar et al. 2007). For consistency with 
the majority of observations, we calculate the faint-end slope for the rest-frame FUV at $z \geq 4$ and at 
lower  redshifts for the rest-frame $B$-band. We indeed find an  evolution in $\alpha$ with redshift that is in fair agreement with the observed evolution. 

\begin{figure}[h]
\centering
	\includegraphics[width=0.45\textwidth]{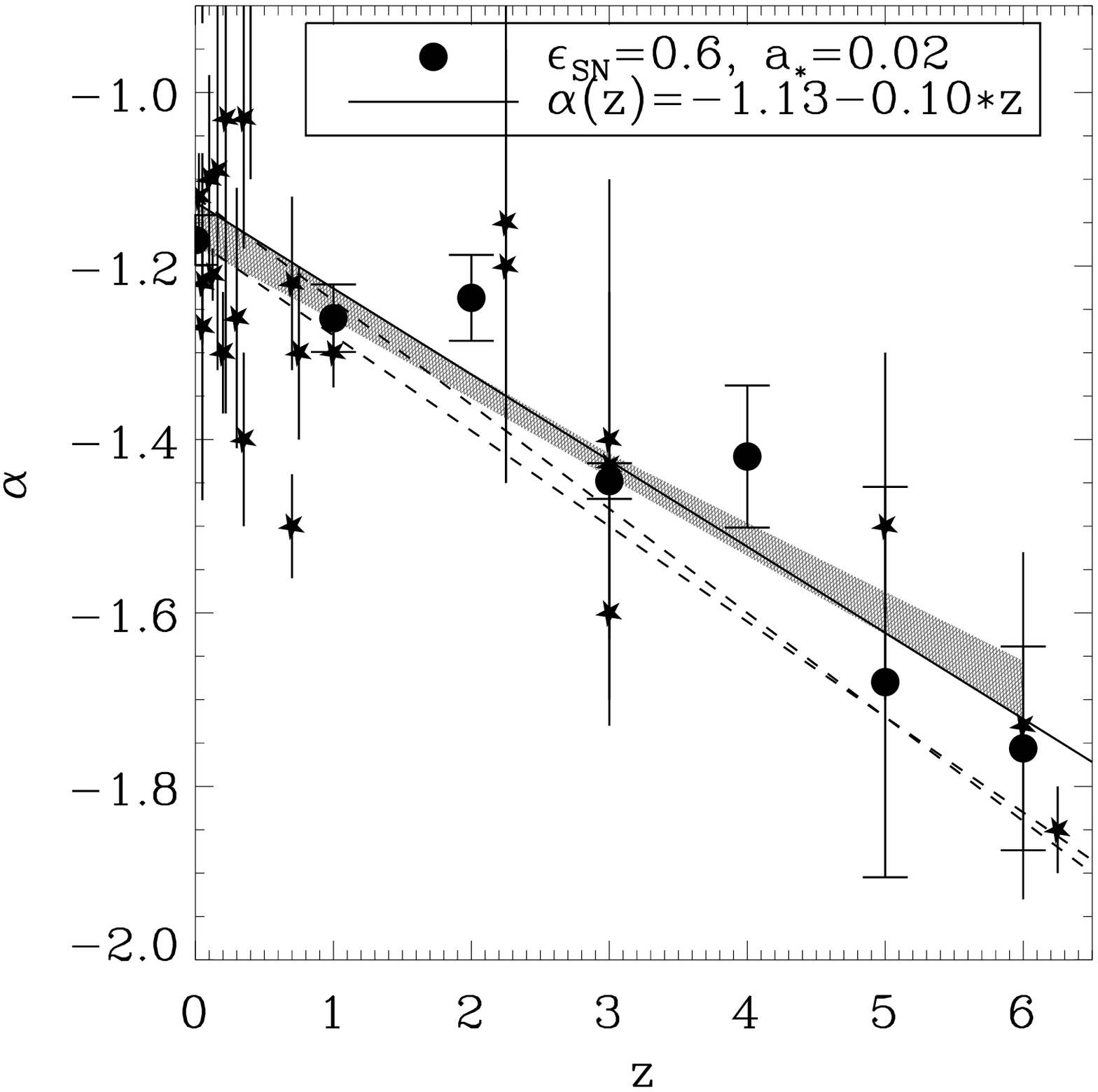} 
\hfill
	\includegraphics[width=0.45\textwidth]{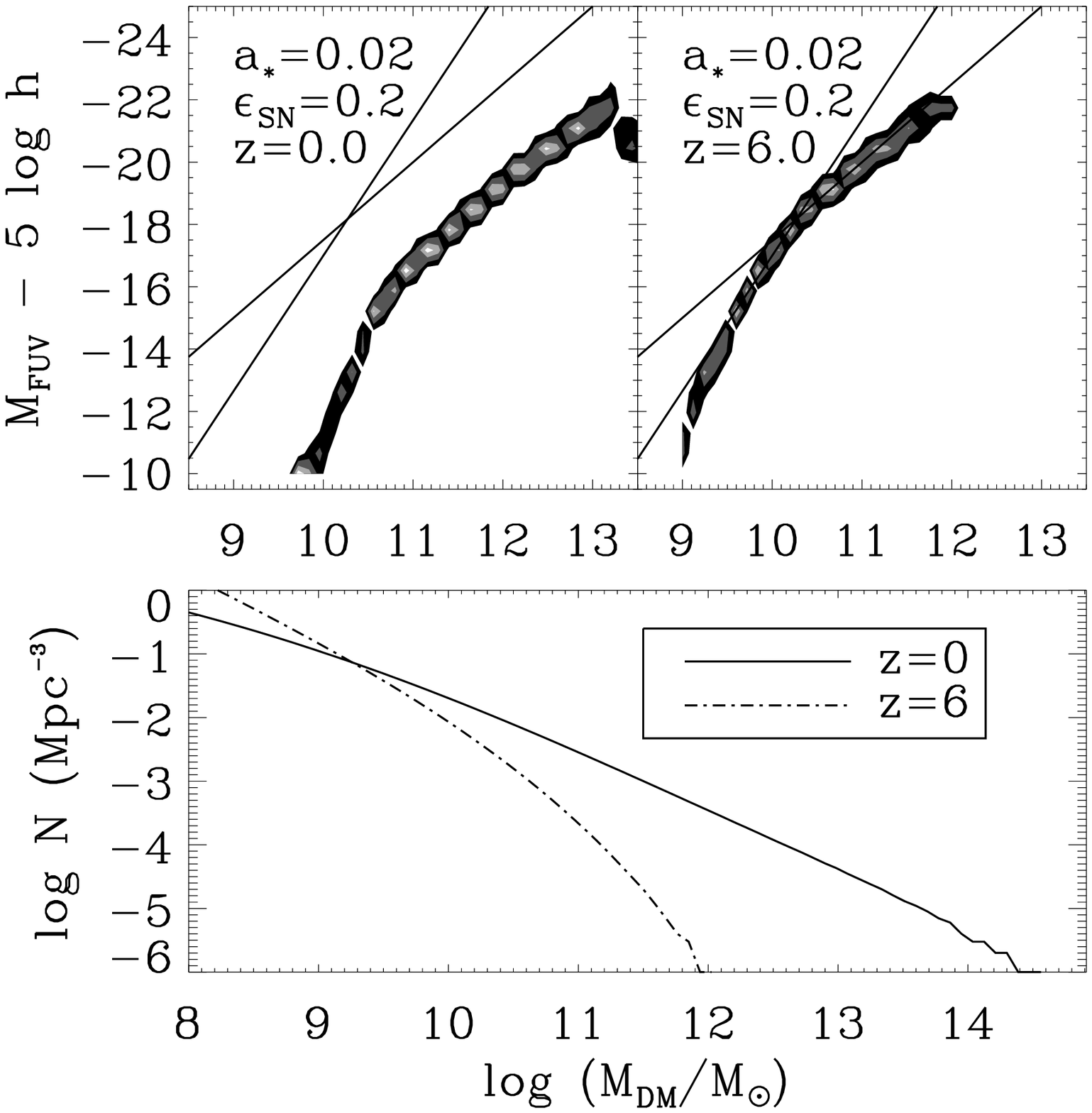} 

  \caption{Left figure: The slope $\alpha$ at different redshifts as predicted by the best fit local model. Filled symbols show results from the simulation and the solid line is the best fit to the simulation data. Errorbars indicate $1-\sigma$ errors. 
    The dashed lines show the fit to the compiled
    data in Ryan et al. (2007). The shaded region shows the range of linear fits to $\alpha(z)$ 
    that we find when varying the star-formation and supernovae feedback efficiency. Stars are the compilation from Ryan et al. (2007). Right figure top panel: Relation between UV luminosity and host dark matter halo mass for central galaxies at $z=0$ and $z=6$. Bottom panel: Dark halo mass function at the same redshifts.}

\end{figure}
The immediate question that arises is, what influences and is the main driver for the evolution
in $\alpha$? Generally, supernova feedback is considered the dominant mechanism in shaping the 
faint-end of the luminosity function {Dekel \& Silk 1986}. The shaded region 
in Fig. 1 shows the range of linear fits to $\alpha(z)$ that we find by varying the star formation efficiency and the supernovae feedback efficiency.  We infer  only a very modest change in $\alpha(z)$ 
for  reasonable choices of feedback efficiencies, and therefore conclude that another process must be responsible for the observed evolution in $\alpha(z)$.   

The mass function of dark matter halos is known to show a strong evolution with redshift. The galaxies contributing to 
the luminosity function around $L_*$ are mostly central galaxies in their dark matter halos, 
i.e. the most luminous galaxy within the halo (e.g. Khochfar \& Ostriker 2008). It is therefore not unreasonable 
to assume a connection between the evolution of $\alpha(z)$ and that of the dark matter mass function.
When considering the luminosity of central galaxies residing in dark matter halos of the same mass
at different redshifts, we find that at early 
times, central galaxies are up to three magnitudes brighter than their counterparts in 
low redshift halos (see Fig. 1 right upper panel). This is even the case 
for halos hosting sub-$L_*$ galaxies. Similar results have been reported by 
(Kobayashi et al. 2007), who showed that dwarf galaxies at early times are not affected 
by supernova feedback in their simulations because cooling times are very short in these halos and 
the energy injected by the supernovae is rapidly   dissipated away.  
The slope in the region of dark matter halos that host sub-$L_*$ galaxies is steeper at high
redshift, and  consequently so is $\alpha$ (Fig. 1 right lower panel). The same is true for other choices supernovae feedback and star formation efficiency, thereby  explaining why we do not find any strong dependence of 
$\alpha(z)$ on these parameters.

\begin{figure}[h]
\centering
	\includegraphics[width=0.45 \textwidth]{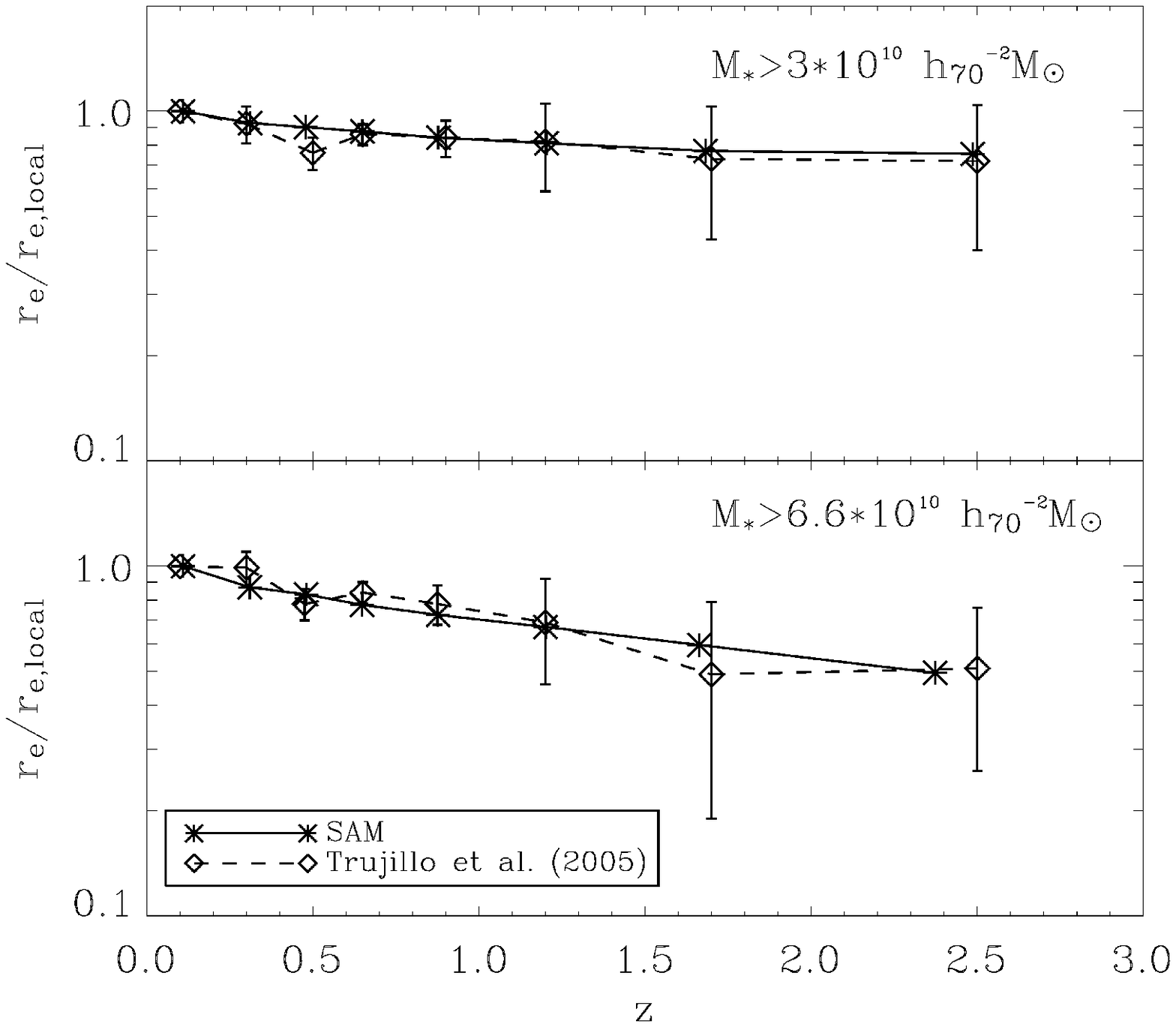} 
\hfill
	\includegraphics[width=0.45 \textwidth]{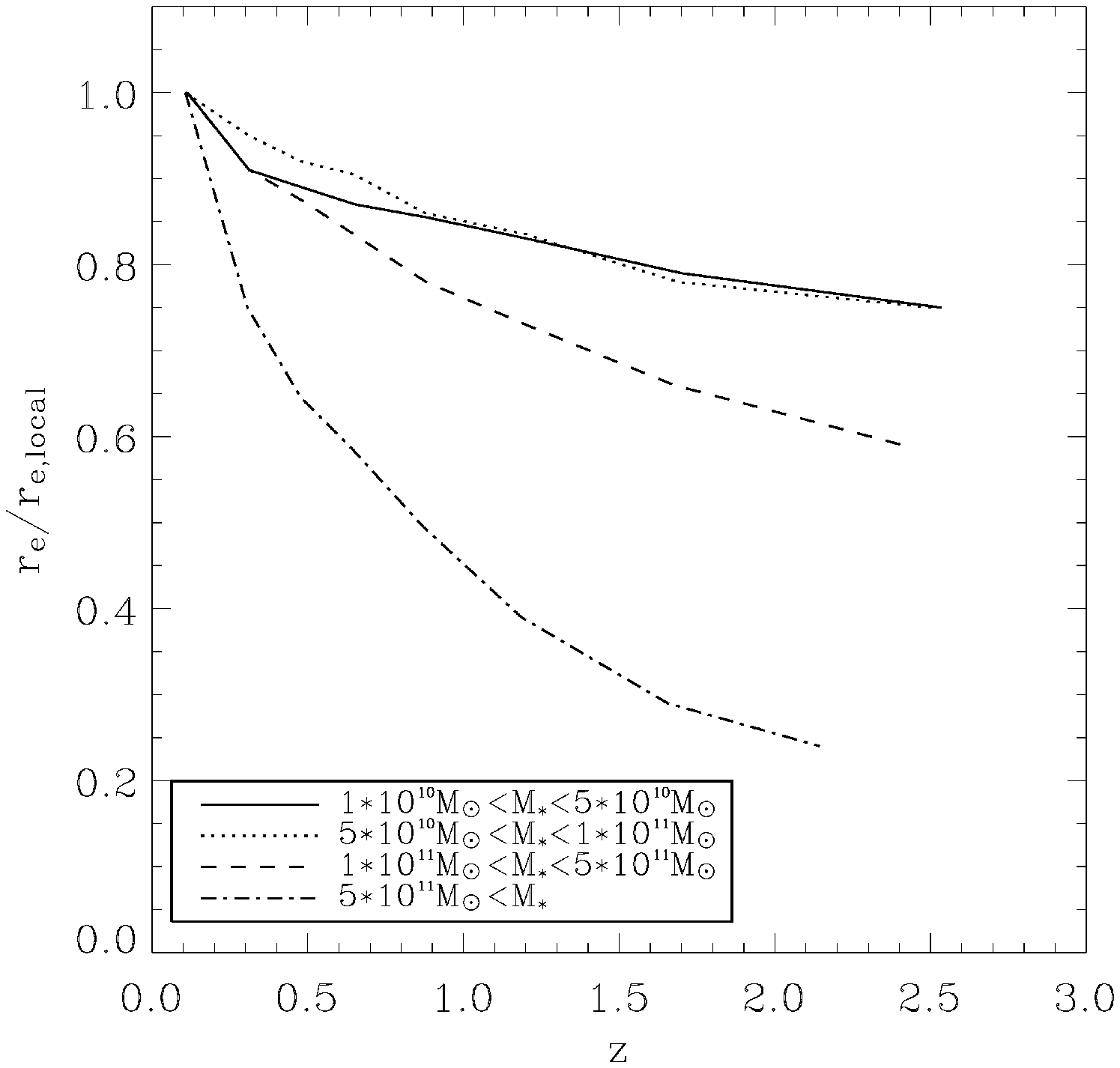} 

  \caption{Left figure: Observed and modeled size-evolution of early-type galaxies of various masses. Right figure: Size-evolution as a function of stellar mass. We find that massive elliptical galaxies show the strongest size-evolution with up to a factor of $ \sim 4$ from $z=2$ to $z=0$.}
\end{figure}

\section{Size-evolution of early-type galaxies}
Various observational surveys have revealed that elliptical galaxies are more compact at high-z (Trujillo et al. 2006). A natural way to explain such a behavior is the combination of dissipation during merger, which drives gas to the centre of the remnant and makes it more compact than in the case of no dissipation (Cox et al. 2006, Khochfar \& Silk 2006b) and the continued accretion of satellites that puff up the host galaxy (Naab et al. 2007, Khochfar \& Silk 2006b). We find that in general the average gas fractions in mergers increase from $10 \%$ in massive mergers at $z=0$ to up to $30 \%$ at $z=4$. In addition to that we find that massive galaxies have an order of magnitude more minor than major mergers. To estimate the size evolution we adopt a model in which the relative size of to remnants of the same mass is proportional the relative amount of dissipation they encountered during their merging history.
In this way a remnant of a gas-rich merger is more compact than that of a gas poor merger and equally a remnant of many gas-poor satellite mergers is less compact than that of a major merger involving gas.

In Fig 2 we calculate the size evolution for the same redshifts presented in 
Trujillo et al. (2006). The authors took the mean effective radii 
of the $\ln(R_{\mbox{e}})$ 
distribution for galaxies above two mass thresholds of 
$3 \times 10^{10} h_{70}^{-2}$ M$_{\odot}$
and $6.6 \times 10^{10} h_{70}^{-2}$ M$_{\odot}$ from the SDDS sample of 
early-type galaxies and divided the effective radii of early-type galaxies at 
higher redshifts by this value. After arranging their galaxies in various 
redshift bins they calculated the means of these ratios and presented 
these values. We here use the same method to compare our results to theirs.
For both cases of limiting 
masses, the agreement is excellent. It appears that the difference in sizes is 
more significant for massive early type galaxies. In the right part of Fig 2 we 
predict the  size-evolution in four different mass ranges based on the 
relative amount of their merger component. While local early-type
galaxies between $10^{10}$ M$_{\odot}$ and $10^{11}$ M$_{\odot}$ are around 
1.25 times larger than their counterparts at $z=2$, 
local  early types with masses larger than $5 \times 10^{11}$ 
M$_{\odot}$ will be approximately 4 times larger than their counterparts at 
$z=2$.  
This dramatic change in sizes in our model results from 
massive galaxies at high redshifts forming 
in  gas-rich mergers  while 
galaxies of the same mass at low redshifts form from dry major mergers (Khochfar \& Burkert 2003, Naab et al. 2006) and minor mergers with small total amounts of cold gas. It is interesting to note that models with a characteristic shut-off mass scale for cooling of gas predict dry mergers as the main mechanism to grow massive galaxies (Khochfar \& Silk 2008a)  and hence imply a large size evolution.

\section{Cold accretion and turbulence}
As mentioned in the introduction, observations reveal a large population of massive, high star forming disk galaxies with gas velocity dispersion of the order $40$ km/s (Genzel et al. 2006). One open question is how these velocity dispersion can be driven. Possible explanations that have been put forward range from gas-rich galaxies mergers (Robertson \& Bullock 2008), tapping into gravitational energy of the disk and supernovae feedback (Bournaud \& Elmegreen 2009, Dekel et al. 2009) to the release of accretion energy by cold streams (Khochfar \& Silk 2008b). We here investigate latter assumption. Numerical simulations by Ocvirk et al. (2008) showed that the fraction of cold accreted gas that makes it way down to 0.2 $R_{vir}$ as a function of dark halo mass is roughly constant down to $z=2$ and a strong declining function of halo mass. We model this behavior by enforcing the simulation results onto the gas cooling rates in our model. This way we make sure to be consistent with the simulation results. Furthermore, we assume that only cold accreted gas is able to deposit a fraction $\eta$ of its kinetic energy once it hits the galactic disk, while hot gas that radiatively cools down does not contribute. Assuming equilibrium between the contributions from accretion and dissipation in the disk leads to a simplified expression for the gas velocity dispersion as 
\begin{eqnarray}
\sigma_g=\sqrt{\eta \sqrt{2} \frac{\dot{M}_{acc}}{M_{gas}}t_{dyn}}V_{vir} 
\end{eqnarray}
with $\dot{M}_{acc}$ the cold accretion rate. We here assumed that the energy dissipation time scale in the galactic disk is proportional to the local dynamical time, and that the contribution to turbulence in the inter-stellar-medium from supernovae is negligible compared to the contribution from accretion, which is the case for high accretion rates that are found  at $z \ge 2$. In the left of Fig 3. we show the predicted correlation between the ratio of disk rotational velocity and gas velocity $V/\sigma$ and disk rotation $V$ from Khochfar \& Silk (2008b). Overlaid are various observations at high redshift showing good agreement between a model in which $eta=0.2$. This result indicates, that only a small fraction is actually necessary to drive sufficiently high turbulence. The strong incline of $V/\sigma$ as a function of $V$ is a direct consequence of the smaller cold accretion fraction in massive halos, thus this model suggest that the functional form of the cold accretion fraction  at high z directly relates to a correlation of $V/\sigma$ and $V$.

\begin{figure}[h]
\centering
	\includegraphics[width=0.7 \textwidth]{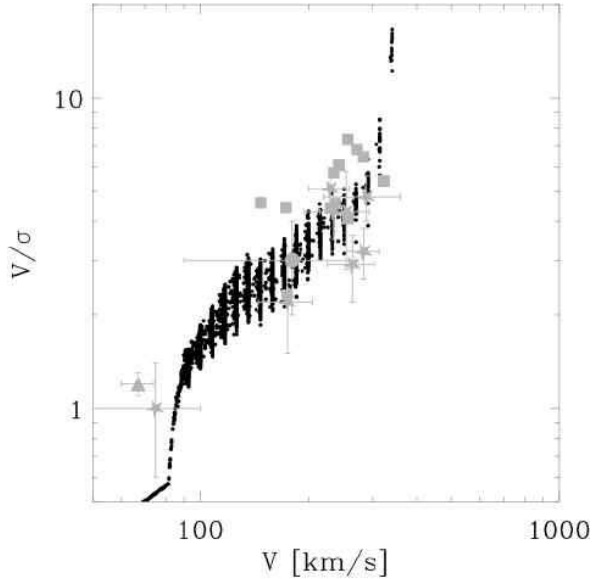} 
  \caption{The relation between $V/\sigma$ and $V$ for modeled galaxies (small filled circles). The observations are from Genzel et al. (2006) (large filled circle), Genzel et al. (2008) (filled stars), Stark et al. (2008) (filled triangle) and Cresci (2008, ApJ submitted) (filled squares). We here assume that $18\%$ of the accretion energy from cold flows is used to drive turbulence in the disk. Note that mergers are excluded as well as satellite galaxies that by construction do not accrete cold gas anymore.}
\end{figure}

\section{Discussion}
In this paper we presented results on the formation of galaxies and their properties at high redshift. We investigated the origin of the evolution of the faint-end luminosity function, finding that in our model it is mainly driven by the evolution of the underlying the dark matter mass function. Low mass galaxies at high z occupy a range in dark matter halos that lies on a much steeper part of the dark matter mass function than their counterparts at low redshifts. This is the main reason for an evolution in the model faint-end slope. Supernovae feedback on the other hand does not introduce an evolving faint-end slope, though it will change the slope to become shallower at the roughly same rate at all redshifts. An interesting possibility in this respect is the choice of IMF. In the case of a top-heavy IMF one would expect more feedback contribution. It has been argued, that the discrepancy between observed star formation rates and integrated stellar mass density in the universe can be reduced by applying a top-heavy IMF at $z \ge 2$. Our simulations however, show that an increase or decrease by a factor of few will not change the general trend of the faint-end slope with redshift, thus not drawing support to the notion of an IMF change at high z.

We continued by focusing on the role that the available fuel for star formation plays. We showed that the formation of galaxies is more efficient at high redshifts, which is a direct result of the fast accretion of cold gas at high z. The accretion of cold gas at high z parallels that of the dark matter, and hence is much faster than at low redshift, resulting in high star formation rates  that can be sustained to build up sufficient massive galaxies. 

Further consequences of high gas fractions in galaxies  at early times are galaxy mergers with large fraction of dissipation. Such mergers have shown to be result in very compact remnants. We applied a simplified model that scales the sizes of remnants with the amount of available fuel during their merger history. Such a model proves to be able to reproduce observed size-evolutions of early-type galaxies very well. We find the strongest size-evolution for the most massive galaxies $ > 5 \times 10^{11}$ M$_{\odot}$, making  a factor of 4 evolution in their  sizes from $z=2$ to $z=0$. The main cause for this strong trend is the occurrence of very gas rich mergers at high z while the local counterparts formed rather recent from dry mergers and a large fraction of satellite galaxies that help to further puff up the host galaxy.

The main mode of accretion at high redshifts is by cold flows of material coming into the halo along with the dark matter halos. In the case of $M_*$ halos at high z these cold accretion flows reach the host halo along cosmic filaments and go straight down to the centre of the halo, where they can feed the main galaxy on a dynamical time. By the time the cold accretion flows reach the main galaxy they will have acquired a substantial amount of kinetic energy that will have to be dispersed. Some if it will be radiated away and some of it will go into rotational energy of the newly arrived material. Another possible option is that some fraction of this energy is used to drive turbulence in the gaseous disk. Observations show that indeed high-z disks tend to have higher gas velocity dispersions than local disks. Using results from numerical simulations on the fraction of cold accreted material as function of halo mass one finds that only $20\%$ of the accretion energy of cold flows is actually needed to drive turbulence at the level that it is observed over a wide range of galaxy masses and rotational velocities of disks at high-z. Interestingly one can recover the steep correlation between $V/\sigma$ and $V$ by just invoking energy equilibrium between accretion energy and energy dissipation in the disk, and the functional dependence of the fraction of cold accretion as a function of halo mass. It will be interesting now to see how the fraction of cold accretion will reflect on $V/\sigma$ for massive galaxies at low z.

We here presented a number of model results on galaxy properties at high redshift, that relied on the faster assembly of structure, in particular the accretion of cold gas onto galaxies. Coming observations will reveal further details of the galaxy population at high redshift, that will certainly require more detailed modeling and will allow to test current models.

\subsection*{References}

{\small

\bref
Binney, J. 1977, ApJ, 215, 483

\bref
Bournaud, F. \& Elmegreen, B. G. 2009, arXiv0902.2806B

\bref
Naab, T., Johansson, P. H., Ostriker, J. P., \& Efstathiou, G. 2007, ApJ, 658, 710

\bref
Dekel, A., \& Silk, J. 1986, ApJ, 303, 39 

\bref
Dekel, A., \& Birnboim, Y. 2006, MNRAS, 368, 2

\bref
Dekel, A.  et al. 2009, arXiv0901.2458D

\bref
F\"orster Schreiber, N. M., et al. 2006, ApJ, 645, 1062

\bref
Genzel, R., et al. 2006, Nature, 442, 786

\bref
Genzel, R., et al. 2008, ApJ, 687, 59 

\bref
Hasinger, G., Miya ji, T., \& Schmidt, M. 2005, A\&A, 441, 417

\bref
Hopkins, A. M. 2004, ApJ, 615, 209

\bref
Juneau, S., et al. 2005, ApJ, 619, L135 

\bref
Keres, D., Katz, N., Weinberg, D. H., \& Dav\'e, R. 2005, MNRAS, 363, 2

\bref
Khochfar, S., \& Burkert, A. 2001, ApJ, 561, 517

\bref
Khochfar, S. \& Burkert, A. 2003, ApJL, 597, L117

\bref
Khochfar, S., \& Burkert, A. 2005, MNRAS, 359, 1379
 
\bref
Khochfar, S., \& Silk, J. 2006a, MNRAS, 370, 902  

\bref
Khochfar, S., \& Silk, J. 2006b, ApJL, 648, L21 

\bref
Khochfar, S., Silk, J., Windhorst, R. A., \& Ryan, R. E., Jr. 2007, ApJL, 668, L115 

\bref
Khochfar, S., \& Ostriker, J. P. 2008, ApJ, 680, 54 

\bref
Khochfar, S., \& Silk, J. 2008a, arXiv0809.1734K (MNRAS submitted)

\bref 
Khochfar, S., \& Silk, J. 2008b, arXiv0812.1183K (ApJL submitted)  
  
\bref
Kobayashi, C., Springel, V., \& White, S. D. M. 2007, MNRAS, 376, 1465 

\bref
Komatsu, E., et al. 2008, arXiv:0803.0547
 
\bref
Le F\`evre O., et al. 2000, MNRAS, 311, 565

\bref
Naab, T., Khochfar, S.,  \& Burkert, A. 2006, ApJ, 636, L81 

\bref
Noeske, K. G., et al. 2007, ApJ, 660, L43 

\bref
Ocvirk, P., Pichon, C., \& Teyssier, R. 2008, MNRAS, 390, 1326

\bref 
P\'erez-Gonz\'alez, P. G. et al. 2008, ApJ, 675, 234

\bref
Robertson, B. E. \& Bullock, J. S. 2008, ApJ, 685, 27

\bref
Rees, M. J., \& Ostriker, J. P. 1977, MNRAS, 179, 541 

\bref
Ryan, R. E., Jr. et al. 2007, ApJ, 668, 839

\bref
Schawinski et al. 2006, Nature, 442, 888

\bref
Stark, D. P., Swinbank, A. M., Ellis, R. S., Dye, S., Smail, I. R., \& Richard, J. 2008, Nature, 
455, 775 

\bref
Somerville, R. S., \& Kolatt, T. S. 1999, MNRAS, 305, 1

\bref
Somerville, R. S. 2002, ApJL, 572, L23

\bref
Springel, V., White, S. D. M., Tormen, G., \& Kauffmann, G. 2001, MNRAS, 328, 726 

\bref
Thomas, D., Maraston, C., Bender, R., \& Mendes de Oliveira, C. 2005, ApJ, 621, 673

\bref
Trujillo, I. et al. 2006, ApJ 650, 18

\bref
Silk, J. 1977, ApJ, 211, 638

\bref
Smith, H.\,D., Miller, P. 1999, A\&A 355, 123 

\bref
your next reference

}

\vfill

\end{document}